\documentclass[letterpaper,onecolumn, 11pt]{IEEEtran}

\usepackage{amsfonts}
\usepackage{cases}
\usepackage{stmaryrd}
\usepackage{color}
\usepackage{amssymb}

\newtheorem{Theorem}{Theorem}
\newtheorem{Corollary}{Corollary}

\newtheorem{Example}{Example}

\newtheorem{Remark}{Remark}
\newtheorem{Lemma}{Lemma}

\newtheorem{Fact}{Fact}

\newcommand{\F}{\mathbb{F}}
\newcommand{\Z}{\mathbb{Z}}
\newcommand{\R}{\mathbb{R}}
\newcommand{\C}{\mathbb{C}}

\begin{document}

%\title{A Tight Upper Bound for the Degree of Bent-Negabent Functions}

\title{Characterization  of Negabent Functions and Construction of Bent-Negabent Functions with Maximum Algebraic Degree}

\date{}

\author{Wei~Su,~
        Alexander~Pott, %~\IEEEmembership{Member,~IEEE,}
        and~Xiaohu~Tang%,~\IEEEmembership{Member,~IEEE,}% <-this stops a space
\thanks{W. Su and X. Tang are with the Institute of Mobile Communications,
Southwest Jiaotong University, Chengdu, 610031, China. Su is
currently a visiting Ph.D. student of the Institute for Algebra and
Geometry (IAG), Otto-von-Guericke University Magdeburg, D-39106
Magdeburg, Germany. (e-mail:  weisu0109@googlemail.com,
xhutang@ieee.org)}% <-this % stops a space
\thanks{A. Pott is with the Institute for Algebra and Geometry (IAG), Otto-von-Guericke University Magdeburg,
D-39106 Magdeburg, Germany. (e-mail: alexander.pott@ovgu.de)}

}

\maketitle

\begin{abstract}
We present necessary and sufficient conditions for a Boolean
function to be a negabent function for both even and odd number of
variables, which demonstrate the relationship between negabent
functions and bent functions. By using these necessary and
sufficient conditions for Boolean functions to be negabent, we
obtain that the nega spectrum of a negabent function has at most 4
values. We determine the nega spectrum distribution of negabent
functions. Further, we provide a method to construct bent-negabent
functions in $n$ variables ($n$ even) of algebraic degree ranging
from $2$ to $\frac{n}{2}$, which implies that the maximum algebraic
degree of an $n$-variable bent-negabent function is equal to
$\frac{n}{2}$. Thus, we answer two open problems proposed by Parker and
Pott and by St\v{a}nic\v{a} \textit{et al.} respectively.
\end{abstract}


\section{Introduction}

Boolean functions play an important role in cryptography and
error-correcting codes. They should satisfy several properties,
which are quite often impossible to be satisfied simultaneously. One
of the most important requirements for Boolean functions is the
nonlinearity, which means that the function is as far away from all
affine functions as possible. In 1976, Rothaus introduced the class
of {\it bent functions} which have the maximum nonlinearity
\cite{Rothaus76}. These functions exist only on even number of
variables and an $n$-variable bent function can have degree at most
$\frac{n}{2}$.

A Boolean function is bent if and only if its spectrum with respect
to the Walsh-Hadamard transform is flat (i.e. all spectral values
have the same absolute value). Parker and Riera extended the concept
of a bent function to some generalized bent criteria for a Boolean
function in \cite{Parker00, Riera06}, where they required that a
Boolean function has flat spectrum with respect to one or more
transforms from a specified set of unitary transforms. The set of
transforms they chose is not arbitrary but is motivated by a choice
of local unitary transforms that are central to the structural
analysis of pure $n$-qubit stabilizer quantum states. The transforms
they applied are $n$-fold tensor products of the identity
$I=\left(\begin{array}{cc}
1 & 0\\
0 & 1
\end{array}\right)$,
the Walsh-Hadamard matrix
$H=\frac{1}{\sqrt{2}}\left(\begin{array}{cc}
1 & 1\\
1 & -1
\end{array}\right)$, and the
nega-Hadamard matrix
$N=\frac{1}{\sqrt{2}}\left(\begin{array}{cc}
1 & i\\
1 & -i
\end{array}\right)$, where $i^{2}=-1$.
The Walsh-Hadamard transform can be described as the tensor product
of several $H's$, and the nega-Hadamard transform is constructed
from the tensor product of several $N's$. As in the case of the
Walsh-Hadamard transform, a Boolean function is called {\it
negabent} if the spectrum under the nega-Hadamard transform is flat.

There are some papers in the area of negabent functions in the last
few years \cite{Parker07}-\cite{Sarkar12}. An interesting topic is
to construct Boolean functions which are both bent and negabent
({\it bent-negabent}), whose relates results are listed as follows.
\begin{enumerate}
\item
In \cite{Parker07}, Parker and Pott gave necessary and sufficient
conditions for quadratic functions to be bent-negabent. It
turns out that such quadratic bent-negabent functions exist for all
even $n$. They also described all Maiorana-McFarland type bent
functions which are simultaneously negabent. It seems difficult to
apply this result in order to construct Maiorana-McFarland
bent-negabent functions. For even number of variables, necessary and sufficient condition for a
Boolean function to be a negabent function
has also been
presented. In \cite{Parker07}, they proposed the following open problem
(open problem 3 in \cite{Parker07}).

%{\it Probelm 1. } Find Boolean functions that are both bent and
%negabent.
%
%{\it Probelm 2. } Determine the number of quadratic bent-negabent functions with $n$ variables ($n$ even).

{\it Open Problem 1: } Find the maximum degree of bent-negabent functions.

\item In \cite{Schmidt08}, transformations that leave the bent-negabent property
invariant are presented. A construction for infinitely many
bent-negabent Boolean functions in $2mn$ variables ($m\not\equiv 1
{\rm \ mod\ 3}$) and of algebraic degree at most $n$ is described,
this being a subclass of the Maiorana-McFarland bent class. Moreover, the algebraic degrees of $n$-variable
bent-negabent functions in this construction are less than or equal
to $\frac{n}{4}$ and $n\equiv 0\ {\rm mod\ }4$. Finally it is shown
that a bent-negabent function in $n$ ($n$ even) variables from the
Maiorona-McFarland class has algebraic degree at most
$\frac{n}{2}-1$, but not an existence result.

\item In \cite{Stanica10}, St\v{a}nic\v{a} \textit{et al.}  developed some properties
of nega-Hadamard transforms. Consequently, they derived several
results on negabentness of concatenations, and partially-symmetric
functions. They also obtained a characterization of bent-negabent
functions in a subclass of Maiorana-McFarland set.

\item In \cite{Stanica12},  St\v{a}nic\v{a} \textit{et al.} pointed out that the
algebraic degree of an $n$-variable  negabent function is at most
$\lceil\frac{n}{2}\rceil$. Further, a characterization of
bent-negabent functions was obtained within a subclass of the
Maiorana-McFarland set.
 They
developed a technique to construct bent-negabent Boolean functions
by using complete mapping polynomials. Using this technique they
demonstrated that for each $l\ge 2$ there exist bent-negabent
functions on $n=12l$ variables with algebraic degree
$\frac{n}{4}+1=l+1$. It is also demonstrated that there exist
bent-negabent functions on 8 variables with algebraic degrees 2, 3
or 4. Moreover, they presented the following open problem.

{\it Open Problem 2: } For any $n\equiv 0{\rm\  mod\ 4}$, give a
general construction of bent-negabent Boolean functions on $n$
variables with algebraic degree strictly greater than
$\frac{n}{4}+1$.

\item In \cite{Sarkar12}, Sarkar considered negabent Boolean functions defined over finite
fields. He characterized negabent quadratic monomial functions. He
also presented necessary and sufficient condition for a
Maiorana-McFarland bent function to be a negabent function. As a
consequence of that result he can obtain bent-negabent
Maiorana-McFarland function of degree $\frac{n}{4}$ over $\F_{2^n}$.

%\item Symmetric Boolean functions form a subclass of Boolean functions.
%A Boolean
%function is called symmetric if the outputs of the function are the
%same for all the inputs with the same weight. In \cite{Savicky94},
%Savicky showed that a symmetric function is bent if and only if it
%is quadratic. In  \cite{Sarkar09}, Sarkar proved that a symmetric
%function is negabent if and only if it is affine.
%This tells that there is no symmetric Boolean function %on even number of variables
%which
%is both bent and negabent.

\end{enumerate}

In this paper, we concentrate on negabent functions and bent-negabent functions.
In particular, we have the following contributions.
\begin{itemize}
\item In Section III,  direct links between the
nega-Hadamard trnsform and the Walsh-Hadamard transform are
explored. By using this property, we study  necessary and sufficient
conditions for a Boolean function to be negabent for both even and
odd number of variables, which demonstrate the relationship between
negabent functions and bent functions.

\item In Section IV,  we obtain
that the nega spectrum of a negabent function has at most 4 values.
Hereafter, we determine the nega spectrum distribution of negabent
functions.

\item In Section V, we give a method to construct
bent-negabent functions in $n$ variables ($n$
even) of %algebraic
degree ranging from $2$ to $\frac{n}{2}$.
%from the Maiorana-McFarland bent functions.
These functions belong to the Maiorana-McFarland complete class.
Thus, we can obtain that the maximum algebraic degree of an
$n$-variable bent-negabent function is equal to $\frac{n}{2}$.
%To
%the best of our knowledge, this is the first ever work on the
%algebraic degree bound of  bent-negabent functions and construction
%of bent-negabent functions with maximum algebraic degree.
Therefore, we answer the Open Problems 1 and 2 proposed in
\cite{Parker07} and \cite{Stanica12} respectively.

\end{itemize}

\section{Preliminaries}

Let $n$ be a positive integer, $\F_2^n$ be the $n$-dimensional
vector space over the two element field $\F_2$. The set of integers,
real numbers and complex numbers are denoted by $\Z$, $\R$ and $\C$,
respectively. To avoid confusion, we denote the addition over $\Z$,
$\R$ and $\C$ by $+$, and the addition over $\F_2^n$ by $\oplus$ for
all $n\ge 1$.

Let $\mathcal{B}_n$ be the set of all maps from $\F_2^n$ to $\F_2$.
Such a map is called an $n$-variable Boolean function. Let
$f(x)\in\mathcal{B}_n$, the {\it support} of  $f(x)$ is defined as
$supp(f)=\{x \in \F_2^n\,|\, f(x) = 1\}$. The {\it Hamming weight}
${\rm wt}(f)$ of $f(x)$ is the size of $supp(f)$, i.e., ${\rm
wt}(f)=|supp(f)|$. The {\it Hamming weight} of a binary vector
$x=(x_1, x_2, \cdots ,x_n)\in \F_2^n$ is defined by ${\rm
wt}(x)=\sum_{i=1}^nx_i$. Each $n$-variable Boolean function $f(x)$
has a unique representation by a multivariate polynomial over
$\F_2$, called the {\it algebraic normal form (ANF)}:
$$f(x_1,\cdots,x_n)=
\bigoplus\limits_{u=(u_1,u_2,\cdots,u_n)\in
\F_2^n}f_u\prod\limits_{i=1}^nx_i^{u_i},\ \ \  f_u\in \F_2.$$ The
algebraic degree, ${\rm deg}(f)$, of $f$ is defined as $\max\{{\rm
wt}(u) | f_u\ne 0, u\in \F_2^n\}$.

The {\em Walsh-Hadamard transform} of $f(x)\in \mathcal{B}_n$ at any
vector $u\in\F_2^n$ is defined by
$$W_f(u)=2^{-\frac{n}{2}}\sum_{x\in \F_2^n}(-1)^{f(x)+u\cdot x},$$
Here $u\cdot x$ is a usual inner product of vectors, i.e., $u\cdot
x=u_1x_1\oplus u_2x_2\oplus\cdots\oplus u_nx_n$ for $u=(u_1, u_2,
\cdots, u_n)$ and $x=(x_1, x_2, \cdots, x_n)\in \F_2^n$. The {\it
Walsh spectrum} of $f$ consists of all values $\{W_f(u)\ |\ u\in
\F_2^n\}$.

 A function $f\in \mathcal{B}_n$ is said to be
{\it bent} if $|W_f(u)|=1$ for all $u\in\F_2^n$. It is {\em
semibent} if
$|W_f(u)|\in \{0, \pm \sqrt{2}\}$. %Note that
Boolean bent (resp.
semibent) functions exist only if the number of variables, $n$, is
even (resp. odd). If $f\in\mathcal{B}_n$ is bent, then the {\it dual function}
$\widetilde{f}$ of $f$, defined on $\F_2^n$ by:
\begin{eqnarray*}
W_f(u)=(-1)^{\widetilde{f}(u)},\ \ \ \forall\ u\in\F_2^n,
\end{eqnarray*} is also bent and its own dual is $f$
itself.

%The sum
%$$C_{f,g}(u)=\sum_{x\in \F_2^n}(-1)^{f(x)\oplus g(x\oplus u)}$$
%is the {\em crosscorrelation} of $f$ and $g$ at $u\in \F_2^n$.
The {\em autocorrelation} of $f$ at $u$ is defined as
$$C_{f}(u)=\sum_{x\in \F_2^n}(-1)^{f(x)\oplus f(x\oplus u)}.$$
For even $n$, it is known that a function $f\in\mathcal{B}_n$ is
bent if and only if $C_f(u)=0$ for all $u\ne (0, 0, \cdots,
0)\in\F_2^n$.

The {\em nega-Hadamard transform} of $f(x)\in \mathcal{B}_n$ at
$u\in\F_2^n$ is the complex valued function:
$$N_f(u)=2^{-\frac{n}{2}}\sum_{x\in \F_2^n}(-1)^{f(x)+u\cdot x}i^{{\rm wt}(x)}.$$
The {\it nega spectrum} of $f$ consists of all values $\{N_f(u)\ |\
u \in \F_2^n\}$.

A function is said to be {\em negabent} if $|N_f(u)|=1$ for all
$u\in\F_2^n$. Note that all the affine functions (both even and odd numbers of
variables) are negabent \cite{Parker07}.
For even number of
variables, if a negabent function is also a bent function, then we
call this function {\it bent-negabent}.

%The sum
%$$c_{f,g}(u)=\sum_{x\in \F_2^n}(-1)^{f(x)\oplus g(x\oplus u)}(-1)^{u\cdot x}$$
%is the {\em nega-crosscorrelation} of $f$ and $g$ at $u\in \F_2^n$.
Define the {\em nega-autocorrelation} of $f$ at $u\in \F_2^n$ by
$$c_{f}(u)=\sum_{x\in \F_2^n}(-1)^{f(x)\oplus f(x\oplus u)}(-1)^{u\cdot x}.$$
In \cite{Stanica10}, it was shown that a Boolean function is negabent if and
only if all its nontrivial nega-autocorrelation values are 0 which is analogous to
the result concerning the autocorrelation values of a bent function.

We conclude this section by introducing the following notations
which will be used throughout this paper.
\begin{enumerate}
\item ${\bf 0}_n=(0, 0, \cdots, 0)$ and ${\bf 1}_n=(1, 1, \cdots, 1)\in \F_2^n$;

\item $e_j$ : $e_j\in\F_2^n$ denotes the
vector of Hamming weight 1 with 1 on the $j$-th component;

\item $\overline{z}$ : if $z=(z_1, \cdots, z_n)\in\F_2^n$, then $\overline{z}=z\oplus {\bf
1}_n$ denotes the bitwise complement of $z$;

%\item $z^*$ : if $z=a+bi\in\C$ is a
%complex number, then $z^*=a-bi$ is the complex
%conjugate of $z$;

\item $|z|$ : if $z=a+bi\in\C$ is a complex number, then $|z|=\sqrt{a^2+b^2}$
denotes the absolute value of $z$;

\item $\sigma_d(x)$ : if $x\in\F_2^n$, then $\sigma_d(x)$
denotes the elementary symmetric Boolean function on $n$ variables
with degree $d$ ($1\le d\le n$), i.e., $$\sigma_d(x)=\bigoplus_{1\le
i_1<\cdots<i_d\le n}x_{i_1}x_{i_2}\cdots x_{i_d},\ \ \ \forall\
x=(x_1,\cdots, x_n)\in\F_2^n.$$ In particular, if $x=(x_1, \cdots,
x_n)\in\F_2^n$, then $\sigma_1(x)=x_1\oplus\cdots \oplus x_n={\bf
1}_n\cdot x$ and $\sigma_2(x)=\bigoplus_{1\le i<j\le n}x_{i}x_{j}$;

\item $GL(n, \F_2)$ : the group of all invertible $n\times n$ matrices over
$\F_2$.
\end{enumerate}

\section{Connections between negabent functions and bent functions}

In this section, direct links between the nega-Hadamard transform
and the Walsh-Hadamard transform are explored. By using this
property, we study necessary and sufficient conditions for a Boolean
function to be negabent for both even and odd number of variables,
which demonstrate the relationship between negabent functions and
bent functions.

\vspace{1mm}

\begin{Lemma}\label{Lem. N_f(u)}
Let $f\in \mathcal{B}_n$. Between the nega-Hadamard transform and
the Walsh-Hadamard transform there is the relation
\begin{eqnarray*}N_f(u)=\frac{W_{f\oplus \sigma_2}(u)+W_{f\oplus \sigma_2}(\overline{u})}{2}
+i\cdot \frac{W_{f\oplus \sigma_2}(u)-W_{f\oplus
\sigma_2}(\overline{u})}{2}.
\end{eqnarray*}
\end{Lemma}

\vspace{1mm}

{\bf Proof}: First for any $x=(x_1, x_2, \cdots, x_n)\in\F_2^n$, it
can be easily proved by induction that
\begin{eqnarray*} {\rm wt}(x)\ ({\rm mod\ }4)%=\sum_{i=1}^nx_i
=\bigoplus_{i=1}^nx_i+2\bigoplus_{1\le i<j\le n}x_ix_j
%\ {\rmmod\ }4
= \sigma_1(x)+2\sigma_2(x)={\bf 1}_n\cdot x+2\sigma_2(x).
\end{eqnarray*}

Thus, the nega-Hadamard transform of $f$ at $u\in\F_2^n$ is
\begin{eqnarray*}
N_f(u)=2^{-\frac{n}{2}}\sum_{x\in \F_2^n}(-1)^{f(x)+u\cdot x}i^{{\rm
wt}(x)}=2^{-\frac{n}{2}}\sum_{x\in
\F_2^n}(-1)^{f(x)+\sigma_2(x)+u\cdot x}i^{{\bf 1}_n\cdot x}.
\end{eqnarray*}
Applying the formula $i^a=\frac{1+(-1)^a}{2}+ i\cdot
\frac{1-(-1)^a}{2}$ for $a\in\F_2$, we get
\begin{eqnarray*}
 N_f(u)&=&2^{-\frac{n}{2}}\sum_{x\in \F_2^n}(-1)^{f(x)+\sigma_2(x)+u\cdot
x}[\frac{1+(-1)^{{\bf 1}_n\cdot x}}{2}+ i\cdot
\frac{1-(-1)^{{\bf 1}_n\cdot x}}{2}]\\
&=&\frac{W_{f\oplus \sigma_2}(u)+W_{f\oplus \sigma_2}(u\oplus {\bf
1}_n)}{2}+ i\cdot \frac{W_{f\oplus \sigma_2}(u)-W_{f\oplus
\sigma_2}(u\oplus
{\bf 1}_n)}{2} \\
&=&\frac{W_{f\oplus \sigma_2}(u)+W_{f\oplus
\sigma_2}(\overline{u})}{2}+ i\cdot \frac{W_{f\oplus
\sigma_2}(u)-W_{f\oplus \sigma_2}(\overline{u})}{2}.
\end{eqnarray*}
\hfill$\Box$

\vspace{1mm}

This property is an important tool to analyse the properties of
negabent functions. If $n$ is even, necessary and sufficient
conditions for a Boolean function $f\in\mathcal{B}_n$ to be negabent
has been given in \cite{Parker07}. By using Lemma \ref{Lem. N_f(u)}
and the Jacobi's two-square theorem, we can obtain the necessary and
sufficient conditions for a Boolean function $f\in\mathcal{B}_n$ to
be negabent for both even and odd $n$. For completeness, we also
provide the proofs for even $n$ here.

\vspace{1mm}

\begin{Fact}(Jacobi's two-square theorem)\label{lem Jacobi}  Let $k$ be a nonnegative integer.
\begin{enumerate}
\item[(1)] The Diophantine equation $x^2+y^2=2^{2k+1}$ has a unique
nonnegative integer solution as $(x,y)=(2^k,2^k)$.
\item[(2)] The Diophantine equation $x^2+y^2=2^{2k}$ has exactly two nonnegative
integer solutions as $(x,y)=(2^k,0)$ and $(x,y)=(0,2^k)$.
\end{enumerate}
\end{Fact}

\vspace{1mm}

\begin{Theorem}(\cite{Parker07})\label{Thm. negabent_even}
Let $n$ be even and $f(x)\in\mathcal{B}_n$. Then $f(x)$ is negabent
if and only if $f(x)\oplus \sigma_2(x)$ is bent.
\end{Theorem}

\vspace{1mm}

{\bf Proof}: A Boolean function $f\in\mathcal{B}_n$ is negabent if
and only if $|N_f(u)|=1$ for all $u\in\F_2^n$. By Lemma \ref{Lem.
N_f(u)}, we have
\begin{eqnarray*}|N_f(u)|^2=
\frac{(W_{f\oplus \sigma_2}(u))^2+(W_{f\oplus
\sigma_2}(\overline{u}))^2}{2}=1,\ \ \forall\ u\in\F_2^n,
\end{eqnarray*}
hence,
\begin{eqnarray*}
(2^{\frac{n}{2}}W_{f\oplus \sigma_2}(u))^2+
(2^{\frac{n}{2}}W_{f\oplus \sigma_2}(\overline{u}))^2=2^{n+1},\ \
\forall\ u\in\F_2^n.
\end{eqnarray*}
From Jacobi's two-square theorem we know that $2^{n+1}$ has a unique
representation as a sum of two squares, namely
$2^{n+1}=(2^{\frac{n}{2}})^2+(2^{\frac{n}{2}})^2$ if $n$ is even.
Thus, it is equivalent to
\begin{eqnarray*}
|2^{\frac{n}{2}}W_{f\oplus \sigma_2}(u)|=|2^{\frac{n}{2}}W_{f\oplus
\sigma_2}(\overline{u})|=2^{\frac{n}{2}},\ \ \forall\ u\in\F_2^n,
\end{eqnarray*}
i.e.,
\begin{eqnarray*}
|W_{f\oplus \sigma_2}(u)|=|W_{f\oplus
\sigma_2}(\overline{u})|=1,\ \ \forall\ u\in\F_2^n.
\end{eqnarray*}
This completes the proof.\hfill$\Box$

\vspace{1mm}

By Theorem \ref{Thm. negabent_even}, the following corollary is
obvious.
%
%\begin{Corollary}\label{Cor. bent-negabent}
%Let $n$ be an even integer and $f\in \mathcal{B}_n$. Then $f$ is
%bent-negabent if and only if both $f$ and $f\oplus \sigma_2$ are
%bent functions.
%\end{Corollary}

\vspace{1mm}

\begin{Corollary}
(\cite{Parker07})\label{Cor. add sigma2} If $f$ is a bent-negabent function,
then $f\oplus \sigma_2$ is also bent-negabent.
\end{Corollary}

\vspace{1mm}

If $n$ is odd, we can get a similar equivalent condition as for even
$n$. In the following, we give three equivalent conditions of a
Boolean function to be negabent for an odd number of variables. The
latter two conditions show the relationship between $n$-variable
negabent functions and $(n-1)$-variable (or $(n+1)$-variable) bent
functions.

\vspace{1mm}

\begin{Theorem}\label{Thm. negabent_odd}
Let $n$ be odd and $f(x)\in\mathcal{B}_n$. Then the following
statements are equivalent:
\begin{enumerate}
\item[(1)]  $f(x)$ is negabent;

\item[(2)]  $f(x)\oplus \sigma_2(x)$ is semibent and
$|W_{f\oplus \sigma_2}(u)|\ne |W_{f\oplus \sigma_2}(\overline{u})|$ for all
$u\in\F_2^n$;

\item[(3)] $(f\oplus \sigma_2)(x_1, \cdots,
x_{n-1}, x_1\oplus x_2 \oplus\cdots\oplus x_n)
=(1\oplus x_n)g(x_1,\cdots, x_{n-1})\oplus x_nh(x_1, \cdots, x_{n-1})$,
where
$g$ and $h$ are both bent functions with $(n-1)$
variables;

\item[(4)] $f(x)\oplus \sigma_2(x)\oplus \sigma_1(x)y$
is bent in $n+1$ variables,
where $x\in\F_2^n$ and $y\in \F_2$.
\end{enumerate}
\end{Theorem}

\vspace{1mm}

{\bf Proof}:  (1) $\Leftrightarrow$ (2): A Boolean function
$f\in\mathcal{B}_n$ is negabent if and only if $|N_f(u)|=1$ for all
$u\in\F_2^n$. It follows from Lemma \ref{Lem. N_f(u)} that
\begin{eqnarray*}|N_f(u)|^2=
\frac{(W_{f\oplus \sigma_2}(u))^2+(W_{f\oplus
\sigma_2}(\overline{u}))^2}{2}=1,\ \ \forall\ u\in\F_2^n,
\end{eqnarray*}
hence,
\begin{eqnarray*}
(2^{\frac{n}{2}}W_{f\oplus \sigma_2}(u))^2+
(2^{\frac{n}{2}}W_{f\oplus \sigma_2}(\overline{u}))^2=2^{n+1},\ \
\forall\ u\in\F_2^n.
\end{eqnarray*}
By Jacobi's two-square theorem, it is
equivalent to
\begin{eqnarray*}
\{|W_{f\oplus \sigma_2}(u)|, |W_{f\oplus
\sigma_2}(\overline{u})|\}=\{0, \sqrt{2}\},\ \ \forall\ u\in\F_2^n.
\end{eqnarray*}
According to the definition of semibent, we can obtain (1) is
equivalent to (2).

(1) $\Leftrightarrow$ (3): Let $f_1(x)=f(x)\oplus \sigma_2(x)$,
$f_2(x)=(f\oplus \sigma_2)(x_1, \cdots, x_{n-1}, x_1\oplus x_2\oplus
\cdots \oplus x_n)$, and the decomposition of $f_2(x)$ is
$f_2(x)=(1\oplus x_n)g(x_1, \cdots, x_{n-1})\oplus x_n h(x_1,
\cdots, x_{n-1})$ for some $g$, $h\in\mathcal{B}_{n-1}$. Then, for
any $v=(v_1,\cdots, v_{n-1}, v_n)\in\F_2^n$, we have
\begin{eqnarray}\label{eqn. W_f2 gh}
W_{f_2}(v)&=&2^{-\frac{n}{2}}\sum_{x'\in\F_2^{n-1},\ x_n\in\F_2}
(-1)^{(1\oplus x_n)g(x')\oplus x_n h(x')\oplus v'\cdot x'\oplus v_nx_n}\nonumber\\
&=&2^{-\frac{n}{2}}\sum_{x'\in\F_2^{n-1}}[ (-1)^{g(x')\oplus v'\cdot
x'}+(-1)^{v_n}
(-1)^{h(x')\oplus v'\cdot x'}]\nonumber\\
&=&\frac{1}{\sqrt{2}}[2^{-\frac{n-1}{2}}\sum_{x'\in\F_2^{n-1}}
(-1)^{g(x')\oplus v'\cdot x'}+(-1)^{v_n}2^{-\frac{n-1}{2}}
\sum_{x'\in\F_2^{n-1}}(-1)^{h(x')\oplus v'\cdot x'}]\nonumber\\
&=&\frac{1}{\sqrt{2}}[W_g(v')+(-1)^{v_n}W_h(v')],
\end{eqnarray}
where $x'=(x_1,\cdots, x_{n-1})$ and $v'=(v_1,\cdots,
v_{n-1})\in\F_2^{n-1}$. Let $\Lambda$ be an $n\times n$ matrix over
$\F_2$ of the form
$$\Lambda=\left(\begin{array}{ccccc}
1 &   &   &   & 1\\
  & 1 &   &   & 1\\
  &   & \ddots&   &\vdots\\
  &   &   & 1 &  1\\
  &   &   &   &1
\end{array}\right),$$
where ``empty" entries are $0$. Then $\Lambda^{-1}=\Lambda$  and
$f_2(x)=f_1(x\Lambda)$. Therefore, for any $v\in\F_2^n$, we can get
that
\begin{eqnarray}\label{eqn. W_f2 f1}
W_{f_2}(v)&=& 2^{-\frac{n}{2}}\sum_{x\in\F_2^n}
(-1)^{f_1(x\Lambda)\oplus v\cdot x}
=2^{-\frac{n}{2}}\sum_{y\in\F_2^n}
(-1)^{f_1(y)\oplus v(y\Lambda)^T}\nonumber\\
&=&2^{-\frac{n}{2}}\sum_{y\in\F_2^n}
(-1)^{f_1(y)\oplus (v\Lambda^T)\cdot y}\nonumber\\
&=&W_{f_1}(v\Lambda^T),
\end{eqnarray}
where the superscript $T$ represents the transpose of a matrix. %Let
%$y=x\Lambda$, then $x=y\Lambda^{-1} =y\Lambda$ and $v\cdot
%x=vx^T=v(y\Lambda)^T=(v\Lambda^T)y^T =(v\Lambda^T)\cdot y$. That is
%the reason we get the second and the third equalities.

For any  $u=(u_1, \cdots, u_{n-1}, u_n)\in\F_2^n$, denote
$w=u\Lambda^T=(w_1, \cdots, w_{n-1}, w_n)\in\F_2^n$.
By equality (\ref{eqn. W_f2 f1}),  we have
$$W_{f_1}(u)=W_{f_2}(u(\Lambda^T)^{-1})=W_{f_2}(u\Lambda^T)=W_{f_2}(w),$$
since $(\Lambda^T)^{-1}=\Lambda^T$. %It follows from
Combined with equality (\ref{eqn. W_f2 gh}), we get
\begin{eqnarray}\label{Eqn_Wf1gh1}W_{f_1}(u)=W_{f_2}(w)=
\frac{1}{\sqrt{2}}[W_g(w')+(-1)^{w_n}W_h(w')], \end{eqnarray} and
\begin{eqnarray}\label{Eqn_Wf1gh2}W_{f_1}(\overline{u})=W_{f_2}((u\oplus {\bf 1}_n)\Lambda^T)
=W_{f_2}(u\Lambda^T\oplus e_n)=W_{f_2}(w\oplus e_n)=
\frac{1}{\sqrt{2}}[W_g(w')-(-1)^{w_n}W_h(w')],\end{eqnarray} where
$w'=(w_1, \cdots, w_{n-1})\in\F_2^{n-1}$. It follows from Lemma
\ref{Lem. N_f(u)}, equalities (\ref{Eqn_Wf1gh1}) and
(\ref{Eqn_Wf1gh2}) that
\begin{eqnarray}\label{Eqn_N_fgh}
N_f(u)&=&\frac{W_{f\oplus \sigma_2}(u)+W_{f\oplus \sigma_2}(\overline{u})}{2}
+i\cdot \frac{W_{f\oplus \sigma_2}(u)-W_{f\oplus
\sigma_2}(\overline{u})}{2}\nonumber\\
&=&\frac{W_{f_1}(u)+W_{f_1}(\overline{u})}{2}
+i\cdot \frac{W_{f_1}(u)-W_{f_1}(\overline{u})}{2}\nonumber\\
&=&\frac{W_g(w')}{\sqrt{2}}+i\cdot (-1)^{w_n}
\frac{W_h(w')}{\sqrt{2}}.
\end{eqnarray}
Since the matrix $\Lambda$ is invertible, we have that
$w=u\Lambda^T=(w', w_n)$ runs over $\F_2^n$ if $u$ runs all over
$\F_2^n$.

Boolean function $f\in\mathcal{B}_n$ is negabent if and only if
$|N_f(u)|=1$ for all $u\in\F_2^n$. It follows from equality
(\ref{Eqn_N_fgh}) that
\begin{center} $|2^{\frac{n-1}{2}}W_{g}(w')|^2+
|2^{\frac{n-1}{2}}W_{h}(w')|^2=2^{n}$,\ \  for all
$w'\in\F_2^{n-1}$.
\end{center}
By Jacobis two-square theorem, it is equivalent to
\begin{center}
$|W_{g}(w')|= |W_{h}(w')|=1$,\ \  for all $w'\in\F_2^{n-1}$,
\end{center}
which means that $g$ and $h$ are both bent functions with $(n-1)$
variables. Therefore, (1) is equivalent to (3).

(2) $\Leftrightarrow$ (4): Let $f'(x, y)=f(x)\oplus
\sigma_2(x)\oplus \sigma_1(x)y=f(x)\oplus \sigma_2(x)\oplus ({\bf
1}_n\cdot x)y\in\mathcal{B}_{n+1}$. Then the Walsh-Hadamard
transform of $f'(x, y)$
at %any vector
$(u, v)\in\F_2^{n+1}$, $u\in\F_2^n$ and $v\in\F_2$, is
\begin{eqnarray}W_{f'}(u, v)&=&2^{-\frac{n+1}{2}}
\sum_{x\in \F_2^n, y\in\F_2}(-1)^{f'(x, y)+u\cdot x+vy}\nonumber\\
&=&2^{-\frac{n+1}{2}}\sum_{x\in \F_2^n}(-1)^{f(x)+
\sigma_2(x)+u\cdot x}+(-1)^v2^{-\frac{n+1}{2}}
\sum_{x\in \F_2^n}(-1)^{f(x)+ \sigma_2(x)+{\bf 1}_n\cdot x+u\cdot x}\nonumber\\
&=&\frac{1}{\sqrt{2}}[W_{f\oplus \sigma_2}(u)+(-1)^vW_{f\oplus
\sigma_2}(\overline{u})].\nonumber
\end{eqnarray}

Then, $f'$ is bent if and only if
$$W_{f'}(u, 0)=\frac{1}{\sqrt{2}}[W_{f\oplus \sigma_2}(u)+W_{f\oplus \sigma_2}(\overline{u})]=\pm 1,
\ \ {\rm for \ all}\ u\in\F_2^n,$$ and
$$W_{f'}(u, 1)=\frac{1}{\sqrt{2}}[W_{f\oplus \sigma_2}(u)-W_{f\oplus \sigma_2}(\overline{u})]=\pm
1, \ \ {\rm for \ all}\ u\in\F_2^n.$$ That is, $|W_{f\oplus
\sigma_2}(u)|\ne |W_{f\oplus \sigma_2}(\overline{u})|$ and
$W_{f\oplus \sigma_2}(u)\in\{0, \pm \sqrt{2}\}$ for all
$u\in\F_2^n$, i.e., $f(x)\oplus \sigma_2(x)$ is semibent.
\hfill$\Box$

\vspace{1mm}

Theorems \ref{Thm.
negabent_even} and \ref{Thm. negabent_odd} demonstrate that negabent
functions and bent functions are closely related.
Theorem \ref{Thm. negabent_odd} also shows that $n$-variable negabent
functions must be semibent if $n$ is odd.

\section{Nega spectrum of negabent functions}

In this section, by using these necessary and sufficient conditions
for Boolean functions to be negabent, we discuss the nega spectrum
distribution of negabent functions.

\vspace{1mm}

\begin{Lemma}\label{Lem. negabent N_f}
Let $f\in\mathcal{B}_n$ be negabent, the values in the nega spectrum
of $f$ are of the form:

\begin{enumerate}
\item[(1)] if $n$ is even, then $N_f(u)\in\{\pm 1,\ \pm i\}$;

\item[(2)] if $n$ is odd, then $N_f(u)\in\{\frac{1+i}{\sqrt{2}},\ \frac{1-i}{\sqrt{2}},\
\frac{-1+i}{\sqrt{2}},\ \frac{-1-i}{\sqrt{2}}\}$.
\end{enumerate}
\end{Lemma}

\vspace{1mm}

{\bf Proof}: (1) If $n$ is even and $f\in\mathcal{B}_n$ is negabent,
then it follows from Theorem \ref{Thm. negabent_even} that $f\oplus
\sigma_2$ is bent. Thus, $W_{f\oplus \sigma_2}(u)=\pm 1$ for all
$u\in\F_2^n$. By Lemma \ref{Lem. N_f(u)}, we have
\begin{eqnarray*}
N_f(u)=\left\{\begin{array}{ll} W_{f\oplus \sigma_2}(u),& {\rm if~\
}
 W_{f\oplus \sigma_2}(u)=W_{f\oplus \sigma_2}(\overline{u}),\\
i\cdot W_{f\oplus \sigma_2}(u),& {\rm if~\ } W_{f\oplus
\sigma_2}(u)\ne W_{f\oplus \sigma_2}(\overline{u}),
\end{array}
\right.
\end{eqnarray*}
for all $u\in\F_2^n$. Therefore, $N_f(u)\in\{\pm 1,\ \pm i\}$.

(2) If $n$ is odd and $f\in\mathcal{B}_n$ is negabent, then it
follows from Theorem \ref{Thm. negabent_odd} that $f(x)\oplus
\sigma_2(x)$ is semibent and $\{|W_{f\oplus \sigma_2}(u)|,\
|W_{f\oplus \sigma_2}(\overline{u})|\}=\{0, \sqrt{2}\}$ for all
$u\in\F_2^n$. By Lemma \ref{Lem. N_f(u)}, we have
$$\begin{array}{c}N_f(u)=\frac{1+i}{2}\cdot W_{f\oplus
\sigma_2}(u)+ \frac{1-i}{2}\cdot W_{f\oplus \sigma_2}(\overline{u}),
\end{array}$$
thus, $N_f(u)\in\{\frac{1+i}{\sqrt{2}},\ \frac{1-i}{\sqrt{2}},\
\frac{-1+i}{\sqrt{2}},\ \frac{-1-i}{\sqrt{2}}\}$.
 \hfill$\Box$

\vspace{1mm}

Lemma \ref{Lem. negabent N_f} shows that the nega spectrum of
negabent function has at most 4 values. This leads to a natural
question of determining the nega spectrum distribution of negabent
functions.

%By using Proposition \ref{Prop. N_f(u)}, we can obtain the following
%result about the size of .

\vspace{1mm}

\begin{Theorem} Let $n$ be even integer and $f\in\mathcal{B}_n$  be
negabent, then the nega spectrum distribution of $f$ is
\begin{eqnarray*}
\left\{\begin{array}{rcc} 1, & 2^{n-2}+2^{\frac{n}{2}-1}&\ {\rm times},\\
-1, & 2^{n-2}-2^{\frac{n}{2}-1}&\ {\rm times},\\
i, & 2^{n-2}&\ {\rm times},\\
-i, & 2^{n-2}&\ {\rm times},
\end{array} \right.\ \
{\rm or}\ \
\left\{\begin{array}{rcc} 1, & 2^{n-2}-2^{\frac{n}{2}-1}&\ {\rm times},\\
-1, & 2^{n-2}+2^{\frac{n}{2}-1}&\ {\rm times},\\
i, & 2^{n-2}&\ {\rm times},\\
-i, & 2^{n-2}&\ {\rm times}.
\end{array} \right.
\end{eqnarray*}
\end{Theorem}

\vspace{1mm}

{\bf Proof}: If $n$ is an even integer and $f\in\mathcal{B}_n$ is
negabent, then by Theorem \ref{Thm. negabent_even}, we have
$f\oplus\sigma_2$ is bent. It is well known that the dual of the
bent function $f\oplus\sigma_2$, $\widetilde{f\oplus\sigma_2}$, is
also bent. By Lemma \ref{Lem. N_f(u)}, %and equation (\ref{Eqn_Dual}),
we can get that
\begin{eqnarray}\label{Eqn_Nfu}
N_f(u)&=&\frac{(-1)^{\widetilde{f\oplus \sigma_2}(u)}+
(-1)^{\widetilde{f\oplus \sigma_2}(\overline{u})}}{2}+i\cdot \frac{
(-1)^{\widetilde{f\oplus \sigma_2}(u)}-(-1)^{\widetilde{f\oplus
\sigma_2}(\overline{u})}}{2}\nonumber\\
&=&\left\{\begin{array}{ll} (-1)^{\widetilde{f\oplus \sigma_2}(u)},
& {\rm if~\ } \widetilde{f\oplus \sigma_2}(\overline{u})=
\widetilde{f\oplus \sigma_2}(u),\\
i\cdot (-1)^{\widetilde{f\oplus \sigma_2}(u)}, &{\rm if~\ }
\widetilde{f\oplus \sigma_2}(\overline{u})\ne \widetilde{f\oplus
\sigma_2}(u),
\end{array}\right.
\end{eqnarray}
for all $u\in\F_2^n$.

For $0\le i,j\le 1$, denote
\begin{eqnarray}\label{Eqn_Sij}
S_{i,j}=|\{u\in\F_2^n|
\widetilde{f\oplus \sigma_2}(u)=i,\ \widetilde{f\oplus
\sigma_2}(\overline{u})=j\}|.
\end{eqnarray}

Recall that  $C_{\widetilde{f\oplus \sigma_2}}(\alpha)=\sum_{u\in \F_2^n}(-1)^{\widetilde{f\oplus \sigma_2}(u)\oplus \widetilde{f\oplus \sigma_2}(u\oplus \alpha)}=0$ for $\alpha\ne {\bf 0}_n$ since $\widetilde{f\oplus \sigma_2}$ is bent, in particular
$$C_{\widetilde{f\oplus \sigma_2}}({\bf 1}_n)=
\sum_{u\in \F_2^n}(-1)^{\widetilde{f\oplus \sigma_2}(u)\oplus
\widetilde{f\oplus \sigma_2}(\overline{u})}=0,$$ which implies
\begin{eqnarray}
S_{0,0}+S_{1,1}&=&2^{n-1},\label{eqn_S0}\\
S_{0,1}+S_{1,0}&=&2^{n-1}.\label{eqn_S1}
\end{eqnarray}
Clearly $S_{1,0}=|\{\overline{u}\in\F_2^n| \widetilde{f\oplus
\sigma_2}(\overline{u})=1,\ \widetilde{f\oplus
\sigma_2}(u)=0\}|=|\{u\in\F_2^n| \widetilde{f\oplus
\sigma_2}(\overline{u})=1,\ \widetilde{f\oplus
\sigma_2}(u)=0\}|=S_{0,1}$. Immediately, it follows from equality
(\ref{eqn_S1}) that $S_{0,1}=S_{1,0}=2^{n-2}$. By equality
(\ref{Eqn_Nfu}),
\begin{eqnarray}\label{eqn. N_f 1}|\{u\in\F_2^n| N_f(u)=i\}|=|\{u\in\F_2^n|
N_f(u)=-i\}|=2^{n-2}.\end{eqnarray}

Since $\widetilde{f\oplus \sigma_2}$ is bent, we have ${\rm
wt}(\widetilde{f\oplus \sigma_2})=2^{n-1}\pm 2^{\frac{n}{2}-1}$. It
is obvious that ${\rm wt}(\widetilde{f\oplus
\sigma_2})=S_{1,0}+S_{1,1}=2^{n-2}+S_{1,1}$. Thus by equality
(\ref{eqn_S0}),
\begin{eqnarray}\label{eqn. 0}
\left\{\begin{array}{c} S_{0,0}=2^{n-2}+2^{\frac{n}{2}-1},\\
S_{1,1}=2^{n-2}-2^{\frac{n}{2}-1},
\end{array} \right.\ \
{\rm or}\ \
\left\{\begin{array}{c} S_{0,0}=2^{n-2}-2^{\frac{n}{2}-1},\\
S_{1,1}=2^{n-2}+2^{\frac{n}{2}-1}.
\end{array} \right.
\end{eqnarray}
Combining equalities (\ref{Eqn_Nfu}), (\ref{Eqn_Sij}), (\ref{eqn.
N_f 1}),
 and (\ref{eqn. 0}),  we get the desired result.
%we can conclude that the
%nega spectrum distribution of $f$ is
%\begin{eqnarray*}
%\left\{\begin{array}{rcc} 1, & 2^{n-2}+2^{\frac{n}{2}-1}&\ {\rm times},\\
%-1, & 2^{n-2}-2^{\frac{n}{2}-1}&\ {\rm times},\\
%i, & 2^{n-2}&\ {\rm times},\\
%-i, & 2^{n-2}&\ {\rm times},
%\end{array} \right.\ \
%{\rm or}\ \
%\left\{\begin{array}{rcc} 1, & 2^{n-2}-2^{\frac{n}{2}-1}&\ {\rm times},\\
%-1, & 2^{n-2}+2^{\frac{n}{2}-1}&\ {\rm times},\\
%i, & 2^{n-2}&\ {\rm times},\\
%-i, & 2^{n-2}&\ {\rm times}.
%\end{array} \right.
%\end{eqnarray*}
%The proof is finished.
\hfill$\Box$

\vspace{1mm}

%\begin{Example}\label{e.g. n=8}
%Take $n=8$ and $f(x)\in\mathcal{B}_8$ is defined as
%$$f(x)=x_1x_2x_3x_4\oplus x_1x_5\oplus x_2x_6\oplus x_3x_7\oplus x_4x_8
%\oplus \sigma_2(x),$$
%where $\sigma_2(x)=\bigoplus_{1\le i<j\le 8}x_ix_j$. Then
%$$f(x)\oplus \sigma_2(x)=x_1x_2x_3x_4\oplus x_1x_5\oplus x_2x_6\oplus x_3x_7\oplus x_4x_8,$$
%is bent function and belongs to Maiorana-McFarland bent class. Therefore, $f$ is negabent.
%Computations show that the nega spectrum distribution of $f$ is
%%The nega-Hadamard spectrum distribution of
%%$f$ is
%\begin{eqnarray*}
%\left\{\begin{array}{rcc} 1, & 72 &\ {\rm times},\\
%-1, & 56 &\ {\rm times},\\
%i, & 64 &\ {\rm times},\\
%-i, & 64 &\ {\rm times}.
%\end{array} \right.
%\end{eqnarray*}
%%{\it  $f$ is not bent.}
%\end{Example}

\begin{Theorem} Let $n$ be odd integer and $f\in\mathcal{B}_n$  be
negabent, then the nega spectrum distribution of $f$ is
\begin{eqnarray*}
\left\{\begin{array}{rcc} \frac{1+i}{\sqrt{2}}, & 2^{n-2}+2^{\frac{n-1}{2}-1}&\ {\rm times},\\
\frac{1-i}{\sqrt{2}}, & 2^{n-2}+2^{\frac{n-1}{2}-1}&\ {\rm times},\\
\frac{-1+i}{\sqrt{2}}, & 2^{n-2}-2^{\frac{n-1}{2}-1}&\ {\rm times},\\
\frac{-1-i}{\sqrt{2}}, & 2^{n-2}-2^{\frac{n-1}{2}-1}&\ {\rm times},
\end{array} \right.\ \
{\rm or}\ \
\left\{\begin{array}{rcc} \frac{1+i}{\sqrt{2}}, & 2^{n-2}-2^{\frac{n-1}{2}-1}&\ {\rm times},\\
\frac{1-i}{\sqrt{2}}, & 2^{n-2}-2^{\frac{n-1}{2}-1}&\ {\rm times},\\
\frac{-1+i}{\sqrt{2}}, & 2^{n-2}+2^{\frac{n-1}{2}-1}&\ {\rm times},\\
\frac{-1-i}{\sqrt{2}}, & 2^{n-2}+2^{\frac{n-1}{2}-1}&\ {\rm times}.
\end{array} \right.
\end{eqnarray*}
\end{Theorem}

{\bf Proof}:  If $n$ is odd and $f\in\mathcal{B}_n$ is negabent,
then by Theorem \ref{Thm. negabent_odd}, we have $$(f\oplus
\sigma_2) (x_1, \cdots, x_{n-1}, x_1\oplus x_2 \oplus\cdots\oplus
x_n) =(1\oplus x_n)g(x_1,\cdots, x_{n-1})\oplus x_nh(x_1, \cdots,
x_{n-1}),$$ where both $g$ and $h$ are bent functions with $(n-1)$
variables. By equality (\ref{Eqn_N_fgh}), we have
\begin{eqnarray}\label{eqn. N_f=a}
N_a=|\{u\in\F_2^n | N_f(u)=a\}|= |\{(w', w_n)\in\F_2^{n-1}\times
\F_2 | \frac{W_g(w')}{\sqrt{2}}+i\cdot (-1)^{w_n}
\frac{W_h(w')}{\sqrt{2}}=a\}|,
\end{eqnarray}
where $a\in\{\frac{1+i}{\sqrt{2}},\ \frac{1-i}{\sqrt{2}},\ \frac{-1+i}{\sqrt{2}},\ \frac{-1-i}{\sqrt{2}}\}$.
%Denote
%\begin{eqnarray*}
%N_{\frac{1+i}{\sqrt{2}}}&=&
%|\{(w', w_n)\in\F_2^{n-1}\times \F_2 |
%\frac{W_g(w')}{\sqrt{2}}+i\cdot (-1)^{w_n}
%\frac{W_h(w')}{\sqrt{2}}=\frac{1+i}{\sqrt{2}}
%\}|,\\
%N_{\frac{1-i}{\sqrt{2}}}&=&
%|\{(w', w_n)\in\F_2^{n-1}\times \F_2 | \frac{W_g(w')}{\sqrt{2}}+i\cdot (-1)^{w_n}
%\frac{W_h(w')}{\sqrt{2}}=\frac{1-i}{\sqrt{2}}\}|,\\
%N_{\frac{-1+i}{\sqrt{2}}}&=&
%|\{(w', w_n)\in\F_2^{n-1}\times \F_2 | \frac{W_g(w')}{\sqrt{2}}+i\cdot (-1)^{w_n}
%\frac{W_h(w')}{\sqrt{2}}=\frac{-1+i}{\sqrt{2}}\}|,\\
%N_{\frac{-1-i}{\sqrt{2}}}&=&
%|\{(w', w_n)\in\F_2^{n-1}\times \F_2 | \frac{W_g(w')}{\sqrt{2}}+i\cdot (-1)^{w_n}
%\frac{W_h(w')}{\sqrt{2}}=\frac{-1-i}{\sqrt{2}}\}|.
%\end{eqnarray*}

Because $g$ is a bent function of $(n-1)$ variables, we have
$|\{w'\in\F_2^{n-1} | W_g(w')=1\}|= 2^{n-2}\pm 2^{\frac{n-1}{2}-1}$.
%\begin{eqnarray*}
%\left\{\begin{array}{lcl}
%|\{w'\in\F_2^{n-1} | W_g(w')=1\}|&=& 2^{n-2}+2^{\frac{n-1}{2}-1},\\
%|\{w'\in\F_2^{n-1} | W_g(w')=-1\}|&=& 2^{n-2}-2^{\frac{n-1}{2}-1},
%\end{array}
%\right.
%\end{eqnarray*}
%or
%\begin{eqnarray*}
%\left\{\begin{array}{lcl}
%|\{w'\in\F_2^{n-1} | W_g(w')=1\}|&=& 2^{n-2}-2^{\frac{n-1}{2}-1},\\
%|\{w'\in\F_2^{n-1} | W_g(w')=-1\}|&=& 2^{n-2}+2^{\frac{n-1}{2}-1},
%\end{array}
%\right.
%\end{eqnarray*}
If $|\{w'\in\F_2^{n-1} | W_g(w')=1\}|= 2^{n-2}+2^{\frac{n-1}{2}-1}$,
then $|\{w'\in\F_2^{n-1} | W_g(w')=-1\}|=
2^{n-2}-2^{\frac{n-1}{2}-1}$. For any $w'\in \{w'\in\F_2^{n-1} |
W_g(w')=1\}$, we can get that
\begin{eqnarray*}
\frac{W_g(w')}{\sqrt{2}}+i\cdot (-1)^{w_n}
\frac{W_h(w')}{\sqrt{2}}=
\left\{\begin{array}{ll}
\frac{1+i\cdot W_h(w')}{\sqrt{2}}, & {\rm if~} w_n=0,\\
\frac{1-i\cdot W_h(w')}{\sqrt{2}}, & {\rm if~} w_n=1,
\end{array}
\right.
\end{eqnarray*}
Since $W_h(w')=\pm 1$ for all $w'\in\F_2^{n-1}$,
we have
$$N_{\frac{1+i}{\sqrt{2}}}=N_{\frac{1-i}{\sqrt{2}}}=2^{n-2}+2^{\frac{n-1}{2}-1}.$$
Because of $|\{w'\in\F_2^{n-1} | W_g(w')=-1\}|= 2^{n-2}-2^{\frac{n-1}{2}-1}$,
we can also get that
$$N_{\frac{-1+i}{\sqrt{2}}}=N_{\frac{-1-i}{\sqrt{2}}}=2^{n-2}-2^{\frac{n-1}{2}-1}.$$
Combining with equality (\ref{eqn. N_f=a}), we  can conclude that
the nega spectrum of $f$ in this case is
\begin{eqnarray*}
\left\{\begin{array}{rcc} \frac{1+i}{\sqrt{2}}, & 2^{n-2}+2^{\frac{n-1}{2}-1}&\ {\rm times},\\
\frac{1-i}{\sqrt{2}}, & 2^{n-2}+2^{\frac{n-1}{2}-1}&\ {\rm times},\\
\frac{-1+i}{\sqrt{2}}, & 2^{n-2}-2^{\frac{n-1}{2}-1}&\ {\rm times},\\
\frac{-1-i}{\sqrt{2}}, & 2^{n-2}-2^{\frac{n-1}{2}-1}&\ {\rm times}.
\end{array} \right.
\end{eqnarray*}

Similarly, if $|\{w'\in\F_2^{n-1} | W_g(w')=1\}|= 2^{n-2}-2^{\frac{n-1}{2}-1}$ and
$|\{w'\in\F_2^{n-1} | W_g(w')=-1\}|= 2^{n-2}+2^{\frac{n-1}{2}-1}$, we can get the nega spectrum of $f$
as follows
\begin{eqnarray*}
\left\{\begin{array}{rcc} \frac{1+i}{\sqrt{2}}, & 2^{n-2}-2^{\frac{n-1}{2}-1}&\ {\rm times},\\
\frac{1-i}{\sqrt{2}}, & 2^{n-2}-2^{\frac{n-1}{2}-1}&\ {\rm times},\\
\frac{-1+i}{\sqrt{2}}, & 2^{n-2}+2^{\frac{n-1}{2}-1}&\ {\rm times},\\
\frac{-1-i}{\sqrt{2}}, & 2^{n-2}+2^{\frac{n-1}{2}-1}&\ {\rm times}.
\end{array} \right.
\end{eqnarray*}
This completes the proof.
\hfill$\Box$

%
%\begin{Example} Take $n=9$, Boolean functions $g(x_1, \cdots, x_8)$,
%$h(x_1, \cdots, x_8)\in\mathcal{B}_8$ are defined as follows
%\begin{eqnarray*}
%g(x_1, \cdots, x_8)&=& x_1x_8\oplus x_2x_7\oplus x_3x_6\oplus
%x_4x_5,\\
%h(x_1, \cdots, x_8)&=& x_5x_6x_7\oplus x_1x_6\oplus x_1x_8 \oplus
%x_2x_5\oplus x_2x_7\oplus x_3x_6\oplus x_4x_5.
%\end{eqnarray*}
%It is easy to check that $g$ and $h$ are both bent functions. Let
%\begin{eqnarray*}
%(f\oplus\sigma_2)(x_1, \cdots, x_8, x_1\oplus x_2\oplus\cdots
%x_9)=(1\oplus x_9)g(x_1, \cdots, x_8)\oplus x_9h(x_1, \cdots, x_8).
%\end{eqnarray*}
%Then we have
%\begin{eqnarray*}
%f(x_1, \cdots, x_8, x_9)=(x_1\oplus x_2\oplus\cdots\oplus
%x_9)(x_1x_6\oplus x_2x_5\oplus x_5x_6x_7)\oplus g(x_1, \cdots ,
%x_8)\oplus \sigma_2(x).
%\end{eqnarray*}
%Computations show that $f$ is negabent and the nega spectrum distribution of $f$ is
%\begin{eqnarray*}
%\left\{\begin{array}{rcc} \frac{1+i}{\sqrt{2}}, & 136 &\ {\rm times},\\
%\frac{1-i}{\sqrt{2}}, & 136 &\ {\rm times},\\
%\frac{-1+i}{\sqrt{2}}, & 120 &\ {\rm times},\\
%\frac{-1-i}{\sqrt{2}}, & 120 &\ {\rm times}.
%\end{array} \right.
%\end{eqnarray*}
%\end{Example}

\section{Construction of Bent-negabent fuctions with maximum algebraic degree}

It is well known that the maximum degree of a bent function on $n$
variables is $\frac{n}{2}$ (for even $n$) \cite{Rothaus76} and the
maximum degree of a negabent function on $n$ variables is
$\lceil\frac{n}{2}\rceil$ (for any integer $n$) \cite{Stanica12}.
But, so far %no such construction is available.
all the known general
constructions of bent-negabent functions on $n$ variables produce
functions with algebraic degrees less than or equal to
$\frac{n}{4}+1$, where  $n$ is any positive integer divisible by $4$
(see \cite{Schmidt08, Stanica12,Sarkar12}).

Throughout this section, let $n=2m$ be any even integer greater than or equal to $4$,
and $h$ be a quadratic bent fucntion defined as $h(x)=\bigoplus_{i=1}^mx_ix_{m+i}$ for
all $x=(x_1, \cdots, x_n)\in\F_2^n$. It is known that any quadratic bent function of $n$ variables
is equivalent to $h(x)$ \cite{Carlet10}.  Since $\sigma_2(x)$ is a quadratic bent
function \cite{Savicky94}, then
there exist $A\in GL(n, \F_2)$, $b$, $u\in\F_2^n$,
and $\epsilon \in \F_2$ such that
\begin{eqnarray}\label{Eqn_Sigma2}
\sigma_2(x)=h(xA \oplus b)\oplus u\cdot x\oplus \epsilon.
\end{eqnarray}
In the sequel, we always assume that $\sigma_2(x)$ is of the above
form as (\ref{Eqn_Sigma2}).

In \cite{Stanica12}, St\v{a}nic\v{a} \textit{et al.}
provided a strategy to construct bent-negabent functions.

\begin{Lemma}\label{Lem. affine eq.}(\cite{Stanica12})
Suppose that both $f\in\mathcal{B}_n$ and  $f\oplus h$  are bent
functions. Then $f'\in\mathcal{B}_n$ defined by
$$f'(x)=f(xA\oplus b)\oplus \sigma_2(x),\ \
x\in\F_2^n,$$ is a bent-negabent function.
\end{Lemma}

\vspace{1mm}

Let $f\in \mathcal{B}_n$ be a Boolean function of the form
\begin{eqnarray}\label{Eqn_MM_Class}
f(x, y)=x\cdot \pi(y)\oplus g(y),\ \ x, y\in\F_2^m,
\end{eqnarray}
where $``\cdot"$ denotes the inner product in $\F_2^{m}$, $\pi :
\F_2^m\shortrightarrow \F_2^m$, and $g : \F_2^m\shortrightarrow
\F_2$. Then the function $f$ is bent if and only if $\pi$ is a
permutation. The whole set of such bent functions forms the
well-known {\it Maiorana-McFarland class}. It is shown in
\cite{Schmidt08} that the degree of a Maiorana-McFarland-type
bent-negabent functions on $n$ variables is at most $\frac{n}{2}-1$
for $n\ge 8$.
%
%
%\begin{Fact}\label{Fact_MM}(\cite{Schmidt08}) The degree of a
%Maiorana-McFarland-type bent-negabent functions on $n$ variables is
%at most $\frac{n}{2}-1$ for $n\ge 8$.
%\end{Fact}

For every positive integer $m$, the vector space $\F_2^m$ can be
endowed with the structure of the finite field $\F_{2^m}$. Any
permutation on $\F_2^m$ can be identified with a permutation of
$\F_{2^m}$. A polynomial $F(X)$ over $\F_{2^m}$ is called a {\it
complete mapping polynomial} if both $F(X)$ and $F(X)+X$ are
permutation polynomials of $\F_{2^m}$. Combining the above Lemma
\ref{Lem. affine eq.} and complete mapping polynomials over
$\F_{2^m}$, St\v{a}nic\v{a} \textit{et al.} gave a method to
construct bent-negabent functions from Maiorana-McFarland bent
functions $f_F(x)=\pi_F(x_1, \cdots, x_m)\cdot (x_{m+1}, \cdots,
x_n)$, where $\pi_F$ denotes the permutation on $\F_2^m$ induced by
a complete mapping polynomial $F(X)\in \F_{2^m}[X]$. However, the
degrees of the bent-negabent functions they constructed are equal to
${\rm deg}(\pi_F)+1$, and there are only few known results on the
complete mapping polynomials with high degrees over $\F_{2^m}$. They
could prove that there exist bent-negabent functions on $n=12l$
variables with algebraic degree $\frac{n}{4}+1=3l+1$, since there
exist complete mapping polynomials on $\F_{2^m}$ of degrees $3l$,
where $m=6l$ and $l\ge 2$ (see \cite{Stanica12, Chapuy07}).

%In fact, if $\pi : \F_2^m\rightarrow\F_2^m$ is a mapping such that
%$\pi(y)$ and $\pi(y)\oplus y$ are permutations, then by using the
%linear transform $A$ and such mapping $\pi$, we can obtain
%$n$-variable bent-negabent functions of degree arranging from $2$ to
%$\frac{n}{2}$ from Maiorana-McFarland bent functions  for any even
%$n\ge 4$. Note that the maximum degree of our bent-negabent
%functions on $n$ variables is equal to $\frac{n}{2}$. Thus, the
%maximum degree of bent-negabent function on $n$ variables is equal
%to $\frac{n}{2}$, and we can obtain bent-negabent functions with
%maximum algebraic degree.

In fact, if $\pi : \F_2^m\rightarrow\F_2^m$ is a mapping such that
$\pi(y)$ and $\pi(y)\oplus y$ are permutations,  from Maiorana-McFarland bent
functions we can construct
infinite class of bent-negabent functions on $n$ variables of degree ranging from $2$ to
$\frac{n}{2}$.  More precisely,  we get the following
results:
\begin{enumerate}

\item We calculate the concrete value of
$A$ in equality  (\ref{Eqn_Sigma2});

\item We show that there exists  mapping $\pi :
\F_2^m\rightarrow\F_2^m$ such that $\pi(y)$ and $\pi(y)\oplus y$ are
permutations and give two methods to get these mappings for any
$m\ge 2$;

\item Using the linear transform $A$ and such mapping $\pi$, we get
bent-negabent functions on $n$ variables of degree arranging from
$2$ to $\frac{n}{2}$ for any even $n\ge 4$.  Note that the maximum
degree of our bent-negabent functions on $n$ variables is equal to
$\frac{n}{2}$. Thus, we answer the Open Problems 1 and 2.
\end{enumerate}

\subsection{The concrete values of $A$, $b$, $u$ and $\epsilon$}

By transforming the quadratic form $\sigma_2$ into its canonical
form, we can obtain that the concrete values of $A=(a_{ij})_{n\times
n}\in GL(n, \F_2)$, $u=(u_1, u_2, \cdots, u_n)$, $b=(b_1, b_2,
\cdots, b_n)\in\F_2^n$, and $\epsilon\in\F_2$ in equality
(\ref{Eqn_Sigma2})
 are
\begin{enumerate}
\item[(1)] $a_{ii}=1$ if $1\le
i\le n$, $a_{ij}=a_{i, m+j}=a_{m+i, j}=a_{m+i, m+j}=1$ if $2\le i\le
m$ and $1\le j\le i-1$, and $a_{ij}=0$ otherwise;

\item[(2)] $u={\bf 0}_n$;

\item[(3)] $b_{2i}=b_{m+2i}=1$ if $1\le i\le \lfloor\frac{m}{2}\rfloor$, and
$b_j=0$ otherwise;

\item[(4)] $\epsilon =1$ if $m\equiv 2, 3 \ ({\rm mod\ }4)$, and $\epsilon=0$
if $m\equiv 0, 1 \ ({\rm mod\ }4)$.
\end{enumerate}

Define matrix $S_m=(s_{ij})_{m\times m}$ over $\F_2$ by
\begin{eqnarray*}
s_{ij}=\left\{
\begin{array}{cc} 1, &\ {\rm if} \ 2\le i\le m,\
1\le j\le i-1;\\
0, &\ {\rm otherwise.}
\end{array}
\right.
\end{eqnarray*}
Then, the $n\times n$ matrix $A$ can be written as
\begin{eqnarray*}
A=\left(
\begin{array}{cc} S_m\oplus I_m & S_m\\
S_m &  S_m\oplus I_m
\end{array}
\right),
\end{eqnarray*}
and $A^{-1}=A$.

\subsection{The existence of mapping $\pi$}

In this subsection, we first explain that there exists  mapping $\pi : \F_2^m\rightarrow\F_2^m$ such that
$\pi(y)$ and $\pi(y)\oplus y$ are permutations for any $m\ge 2$ from the
perspective of the complete mapping polynomial over finite field $\F_{2^m}$. And then
introduce two methods to obtain the mapping $\pi$
directly from the vector space $\F_2^m$.

If $\sigma(x)$ is a complete mapping polynomial over $\F_{2^m}$,
then the corresponding permutation $\sigma'(x)$ on  $\F_2^m$
satisfies $\sigma'(x)$ and $\sigma'(x)\oplus x$ are both
permutations. Trivial examples of complete mapping polynomials are
the linear polynomials $\sigma(x)=ax$ with $a\ne 0$, $-1$. If $m\ge
3$, there exist complete mapping polynomials of $\F_{2^m}$ of
reduced degree $>1$. For details on complete mapping polynomials we
refer to \cite{Niederreiter82}. Thus, there exists  mapping $\pi :
\F_2^m\rightarrow\F_2^m$ such that both $\pi(y)$ and $\pi(y)\oplus
y$ are permutations for any $m\ge 2$.
%
%the mapping $\pi : \F_2^m\rightarrow\F_2^m$ given in Theorem
%\ref{Thm. bent-negabent MM} exists.

In what follows, we introduce two methods to obtain the linear permutation $\pi : \F_2^m\rightarrow\F_2^m$ such that $\pi(y)\oplus y$
is also permutation for any $m\ge 2$.
Define the mapping $\pi :\F_2^m \shortrightarrow \F_2^m$ as
$\pi(y)=yM$, where $y=(y_1, y_2, \cdots, y_m)\in\F_2^m$.  If we can find $m\times m$ matrix $M$ over $\F_2$ such that $M$ and
$M\oplus I_m$ have full rank $m$, then we get the desired linear permutation $\pi$.

If $m=2$, there are two matrices satisfy the conditions:
\begin{eqnarray*}\left(\begin{array}{cc}
1 & 1\\
1 & 0
\end{array}\right)\ \ {\rm and}\ \
\left(\begin{array}{cc}
0 & 1\\
1 & 1
\end{array}\right).
\end{eqnarray*}
Using exhaustive computer search, we found that there are 48
matrices satisfying the conditions for $m=3$, and 5824 matrices
satisfying the conditions for $m=4$. For example,
\begin{eqnarray*}\left(\begin{array}{ccc}
0 & 1 & 1\\
1 & 1 & 0\\
1 & 0 & 0
\end{array}\right),\ \ \ \
\left(\begin{array}{ccc}
1 & 1 & 1\\
0 & 1 & 1\\
1 & 0 & 1
\end{array}\right),
\end{eqnarray*}
and \begin{eqnarray*}
\left(\begin{array}{cccc}
0 & 1 & 0 & 1\\
1 & 0 & 1 & 0\\
0 & 1 & 0 & 0\\
1 & 0 & 0 & 0
\end{array}\right),\ \  \ \
\left(\begin{array}{cccc}
1 & 0 & 1 & 1\\
0 & 1 & 1 & 0\\
1 & 1 & 0 & 0\\
1 & 0 & 0 & 0
\end{array}\right).
\end{eqnarray*}

%For $m\ge 4$, we have the following two methods to get the square matrix
%$M$, which satisfies that $M$ and $M\oplus I_m$ have full rank $m$.

{\bf Method 1. } For any  even $m\ge 4$, Parker and Pott gave a
method to construct $m\times m$ symmetric matrix $M$ over $\F_2$
such that $M$ and $M\oplus I_m$ have rank $m$ in Section 3 of
\cite{Parker07}. To save space, here we will not give the detail.

{\bf Method 2. }  An  $m\times m$ block matrix $P$ is said to be
{\it block diagonal matrix} if it has main diagonal blocks square
matrices such that the off-diagonal blocks are zero matrices, i.e.,
$P$ has the form
\begin{eqnarray*}
P = \left( \begin{array}{cccc}
P_1 & 0 & \cdots & 0\\
0 & P_2 & \cdots & 0\\
\vdots  & \vdots & \ddots & \vdots \\
0 &  0 & \cdots & P_t
\end{array}\right),
\end{eqnarray*}
where $P_j$, $1\le j\le t$, is a square matrix of order $k_j$, and
$k_1+\cdots +k_t=m$. It can be indicated as ${\rm diag}(P_1, P_2,
\cdots, P_t)$. Any square matrix can trivially be considered a block
diagonal matrix with only one block.

For the determinant of block diagonal matrix $P$, the following
property holds
$${\rm det}(P)=\prod_{i=1}^t{\rm det}(P_i).$$
By this property of diagonal matrix, we can easily get the following
recursive construction.

\begin{Lemma}\label{Lem. matrix M} Let $t\ge 2$ and $M_j$
 be a square matrix of order $k_j$
 such that $M_j$ and $M_j\oplus I_{k_j}$ have full rank for any $1\le j\le
 t$. If $k_1+\cdots +k_t=m$,
then the matrix $M={\rm diag}(M_1, M_2, \cdots, M_t)$
%\begin{eqnarray*}
%M = \left( \begin{array}{cccc}
%M_1 & 0 & \cdots & 0\\
%0 & M_2 & \cdots & 0\\
%\vdots  & \vdots & \ddots & \vdots \\
%0 &  0 & \cdots & M_t
%\end{array}\right)
%\end{eqnarray*}
and $M\oplus I_m$ have rank $m$.
\end{Lemma}

As mentioned before, for $m=2$, $3$, there exists matrix $M$ such
that both $M$ and $M\oplus I_m$ have full rank. Thus, for any $m\ge
2$, we can get matrix $M$ such that $M$ and $M\oplus I_m$ have full
rank by Lemma \ref{Lem. matrix M}.
Therefore, the linear permutation $\pi(y)=yM$ has been obtained.

\subsection{Construction for
infinite class of bent-negabent functions}

%Several transformations that preserve the bent-negabent property
%have been presented in \cite{Parker07, Schmidt08}.
%Here we introduce
%some transformations.

If $f\in \mathcal{B}_n$ is a bent function, then the function given
by
\begin{eqnarray}\label{Eqn_Complete_Class}
f(x\cdot C\oplus \alpha)\oplus \beta \cdot x\oplus \zeta,\ \ {\rm
where}\ C\in GL(n, \F_2),\ \alpha, \beta\in\F_2^n,\ \zeta\in\F_2,
\end{eqnarray}
 is also bent. All the  functions in (\ref{Eqn_Complete_Class}) is called a {\it complete class}.
 Specifically, it is said to be Maiorana-McFarland complete class
 if $f$ belongs to  Maiorana-McFarland class
in (\ref{Eqn_MM_Class}).

Counterexamples show that these operations generally do not preserve
the negabent property of a Boolean function. %\cite{}.
Indeed if $GL(n, \F_2)$ is replaced by $O(n, \F_2)$, the orthogonal
group of $n\times n$ matrices over $\F_2$, the negabent property is
still preserved.

\begin{Lemma}(\cite{Schmidt08})\label{Lem. add affine}
Let $f$, $g : \F_2^n \rightarrow \F_2$ be two Boolean functions.
Suppose that $f$ and $g$ are related by $g(x)=f(x\cdot O\oplus
\alpha)\oplus \beta \cdot x\oplus \zeta$, where $O$ is an $n\times
n$ orthogonal matrix over $\F_2$, $\alpha$, $\beta\in\F_2^n$, and
$\zeta\in\F_2$. Then, if $f$ is bent-negabent, $g$ is also
bent-negabent.
\end{Lemma}

Now, we are ready to
construct $2m$-variable bent-negabent functions of  degree ranging
from $2$ to $m$.

\vspace{1mm}

\begin{Theorem}\label{Thm. bent-negabent MM}
Define $f\in \mathcal{B}_n$ by
$$f(x, y)=x\cdot \pi(y)\oplus g(y),\ \ x, y\in\F_2^m,$$
where $\pi :\F_2^m \shortrightarrow \F_2^m$ is a mapping such that
$\pi(y)$ and $\pi(y)\oplus y$ are permutations and
$g\in\mathcal{B}_m$. Then
\begin{eqnarray}\label{Eqn_bent-neg-max}
f'(x, y)=f((x, y)\cdot OA\oplus \alpha)\oplus \beta\cdot x\oplus \zeta
\end{eqnarray}
is a bent-negabent function with $\mathrm{deg}(f')=\mathrm{deg}(f)$,
for any $\alpha$, $\beta\in\F_2^n$, $\zeta\in\F_2$, and any $n\times
n$ orthogonal matrix $O$ over $\F_2$.
\end{Theorem}

{\bf Proof}: If $\pi(y)$ and $\pi(y)\oplus y$ are permutations on
$\F_2^m$, we have that $f(x,y)$ and $f(x,y)\oplus
h(x,y)=f(x,y)\oplus x\cdot y$ are both Maiorana-McFarland bent
functions. It follows from Lemma \ref{Lem. affine eq.} and Corollary
\ref{Cor. add sigma2}   that $f((x,y)\cdot A\oplus b)$ is a bent-negabent
function. Applying Lemma \ref{Lem. add affine} to $f((x,y)\cdot A\oplus
b)$, we have that $f((x, y)\cdot OA\oplus \alpha)\oplus \beta\cdot
x\oplus\zeta$ is also a bent-negabent function for any $\alpha$,
$\beta\in\F_2^n$, $\zeta\in\F_2$, and any $n\times n$ orthogonal
matrix $O$ over $\F_2$.

Since the algebraic degree is an affine invariant, we have
$\mathrm{deg}(f')=\mathrm{deg}(f)$.
\hfill$\Box$

Note that we are free to choose $g$. Specifically if taking $g\in\mathcal{B}_m$ with $\mathrm{deg}(g)=m$, one has
$\mathrm{deg}(f')=\mathrm{deg}(f)=m$. %Then we can obtain
%$n$-variable bent-negabent functions of maximum degrees.
%Therefore, the following result holds.
It is well known that the maximum degree of bent function in $2m$
variables is $m$. Then, the maximum degree of bent-negabent function
in $2m$ variables is less than or equal to $m$. Our construction can
reach the maximal degree, so the bound is tight. Therefore, the
following result holds.

%Thus, the degrees of the
%bent-negabent functions constructed in Theorem \ref{Thm.
%bent-negabent MM} can reach $\frac{n}{2}$. 

%\vspace{1mm}
%
%\begin{Corollary}\label{Cor_deg=m_1}
%Define $f\in \mathcal{B}_n$ by
%$$f(x, y)=x\cdot \pi(y)\oplus g(y),\ \ x, y\in\F_2^m,$$
%where $\pi :\F_2^m \shortrightarrow \F_2^m$ is a mapping such that
%$\pi(y)$ and $\pi(y)\oplus y$ are permutations, and
%$g\in\mathcal{B}_m$ with $\mathrm{deg}(g)=m$. Then $f'(x, y)=f((x,
%y)OA\oplus \alpha)\oplus \beta\cdot x\oplus \zeta$ is a
%bent-negabent function with $\mathrm{deg}(f')=m$, for any $\alpha$,
%$\beta\in\F_2^n$, $\zeta\in\F_2$, and any $n\times n$ orthogonal
%matrix $O$ over $\F_2$.
%\end{Corollary}

\vspace{1mm}

\begin{Corollary}\label{Cor. bent-negabent degree}
Let $n$ be even and $f\in\mathcal{B}_n$. If $f$ is bent-negabent,
then the algebraic degree of $f$ is at most $\frac{n}{2}$. And the
bent-negabent function $f'$ given by (\ref{Eqn_bent-neg-max}) can
achieve the maximal algebraic degree if $\mathrm{deg}(g)=m$ or ${\rm
deg}(\pi)=m-1$.
\end{Corollary}

\vspace{1mm}

\begin{Remark}
Since the degree of a
Maiorana-McFarland-type bent-negabent function on $n$ variables is
at most $\frac{n}{2}-1$ for $n\ge 8$ (see \cite{Schmidt08}), the
functions constructed by Theorem \ref{Thm. bent-negabent MM} may not in the Maiorana-McFarland class, but belong to the
Maiorana-McFarland complete class.
\end{Remark}

The dual also preserve the bent-negabent function property.

%
%\begin{Lemma}
%\label{Lem. affine trans.}
%If $f$, $f\oplus
%h\in\mathcal{B}_n$ are both bent functions. Then $f'(x)=f(xOA\oplus
%\alpha)\oplus \beta\cdot x\oplus \zeta$ is a bent-negabent function
%for any $\alpha$, $\beta\in\F_2^n$, $\zeta\in\F_2$, and any $n\times
%n$ orthogonal matrix $O$ over $\F_2$.
%\end{Lemma}
%
%{\bf Proof}: It follows from Lemmas \ref{Lem. affine eq.} and
%\ref{Cor. add sigma2} that $g'(x)=f(xA\oplus b)$ is a bent-negabent
%function. By Lemma \ref{Lem. add affine}, we can obtain
%that
%$$g'(xO+\alpha')\oplus \beta\cdot x\oplus \zeta=
%f(xOA\oplus \alpha'A\oplus b)\oplus \beta\cdot x\oplus \zeta$$ is a
%bent-negabent function for any $\alpha'$, $\beta\in\F_2^n$, and
%$\zeta\in\F_2$. Since $A$ is invertible, we have
%$\alpha=\alpha'A\oplus b$ runs over $\F_2^n$ if $\alpha'$ runs all
%over $\F_2^n$. Thus,
% $ f(xOA\oplus \alpha)\oplus \beta\cdot x\oplus \zeta$ is a
%bent-negabent function for any $\alpha$, $\beta\in\F_2^n$,
%$\zeta\in\F_2$, and any $n\times
%n$ orthogonal matrix $O$ over $\F_2$. \hfill$\Box$

%\subsection{Construction for
%infinite class of bent-negabent functions over $\F_2^{2m}$ with maximum algebraic degree $m$}

\begin{Lemma}(\cite{Parker07})\label{Lem_dual}
If $f$ is a bent-negabent function, then its dual is again
bent-negabent.
\end{Lemma}

\begin{Lemma}\label{Lem_deg_dual}(\cite{Carlet10})
The algebraic degrees of any $n$-variable bent function $f$ and
of its dual $\widetilde{f}$ satisfy:
$$\begin{array}{c}
\frac{n}{2}-{\rm deg}(f)\ge \frac{\frac{n}{2}-
{\rm deg}(\widetilde{f})}{{\rm deg}(\widetilde{f})-1}.
\end{array}$$
\end{Lemma}

It follows from Lemma \ref{Lem_deg_dual} that the degree of $\widetilde{f}$, ${\rm deg}(\widetilde{f})$, is also equal to $\frac{n}{2}$ if $f$
is an $n$-variable bent function with ${\rm deg}(f)=\frac{n}{2}$. Combining
Lemma \ref{Lem_dual} and Lemma \ref{Lem_deg_dual}, we have the following corollary.

\begin{Corollary}\label{Cor_deg=m_2}
Let $p(x)\in\mathcal{B}_n$ be a bent-negabent function with degree
$m$ obtained from Theorem \ref{Thm. bent-negabent MM}. Then its dual
is again bent-negabent with degree $m$.
\end{Corollary}

%
%
%
%In the reminder of this paper, we first calculate the concrete values of
%$A$, $b$, $u$, and $\epsilon$. And then introduce two methods to obtain the mapping $\pi :
%\F_2^m\rightarrow\F_2^m$ such that $\pi(y)$ and $\pi(y)\oplus y$ are
%permutations. Finally, we give two examples of bent-negabent functions in $8$
%and $10$ variables with degree $4$ and $5$, respectively.

\subsection{Examples of $n$-variable bent-negabent functions with maximum degree for $n=8$
and $n=10$ }

\begin{Example} Take $m=4$, $n=2m=8$, $\pi(y)=yM$ with matrix
\begin{eqnarray*}
M=\left(
\begin{array}{cccc}
 0 & 1 & 0 & 1\\
 1 & 0 & 1 & 0\\
 0 & 1 & 0 & 0\\
 1 & 0 & 0 & 0
\end{array}\right),
\end{eqnarray*}
and $g(y)=y_1y_2y_3y_4$ in Theorem \ref{Thm. bent-negabent MM}. It
is easy to check that matrices $M$ and $M\oplus I_4$ have rank $4$.
Then
$$\pi(y)=yM=(y_2\oplus y_4,\ y_1\oplus y_3,\ y_2,\ y_1),$$
and
$$f(x, y)=x\cdot \pi(y)\oplus g(y)=
x_1\cdot(y_2\oplus y_4)\oplus
x_2\cdot( y_1\oplus y_3)\oplus x_3\cdot y_2\oplus x_4\cdot y_1\oplus
y_1y_2y_3y_4.$$
The linear transformation matrix $A$ is equal to
\begin{eqnarray*}
A=\left(
\begin{array}{cc}
S_4\oplus I_4 & S_4 \\
S_4 & S_4\oplus I_4 \\
\end{array}\right),\ \ {\rm where}
\ \
S_4=\left(
\begin{array}{cccc}
 0 & 0 & 0 & 0\\
 1 & 0 & 0 & 0\\
 1 & 1 & 0 & 0\\
 1 & 1 & 1 & 0
\end{array}\right).
\end{eqnarray*}
Let $O=I_n$, $\alpha=\beta={\bf 0}_n$ and $\zeta=0$.
Then the function $ f'(x, y)=f((x, y)A)= x_2x_3x_4y_4\oplus
x_2x_3y_3y_4\oplus x_2x_4y_2y_4\oplus x_2y_2y_3y_4\oplus
x_3x_4y_1y_4\oplus x_3y_1y_3y_4\oplus x_4y_1y_2y_4\oplus
y_1y_2y_3y_4 \oplus x_2x_3y_4\oplus x_2x_4y_4\oplus x_2y_2y_4\oplus
x_2y_3y_4 \oplus x_3x_4y_4\oplus x_3y_1y_4\oplus x_3y_3y_4 \oplus
x_4y_1y_4\oplus x_4y_2y_4\oplus y_1y_2y_4 \oplus y_1y_3y_4\oplus
y_2y_3y_4 \oplus x_1x_3\oplus x_1x_4 \oplus x_1y_2\oplus x_1y_3
\oplus x_2x_3\oplus x_2x_4 \oplus x_2y_1\oplus x_3y_1 \oplus
x_3y_4\oplus x_4y_2 \oplus x_4y_4 \oplus y_1y_3 \oplus y_2y_3 \oplus
y_3y_4 \oplus x_2\oplus x_3\oplus x_4 \oplus y_2 \oplus y_3$ is bent-negabent 
and $\mathrm{deg}(f')=4$.
%
%Computations show that
%$f'$ is bent-negabent.
\end{Example}

\vspace{1mm}

\begin{Example} Take $m=5$, $n=2m=10$, $\pi(y)=yM$ with matrix
\begin{eqnarray*}
M=\left(
\begin{array}{cc}
 M_1 & 0\\
 0 & M_2
\end{array}\right),\ \
{\rm where}\ \
M_1=\left(
\begin{array}{cc}
 1 & 1\\
 1 & 0
\end{array}\right),
 \ \ {\rm and}\ \
M_2=\left(
\begin{array}{ccc}
 0 & 1 & 1\\
 1 & 1 & 0\\
 1 & 0 & 0
\end{array}\right),
\end{eqnarray*}
and $g(y)=y_1y_2y_3y_4y_5\oplus y_2y_3y_4y_5$. It is easy to check
that matrices $M$ and $M\oplus I_5$ have rank $5$. Then
$$\pi(y)=yM=(y_1\oplus y_2,\ y_1,\ y_4\oplus y_5,\ y_3\oplus y_4,\ y_3),$$
and
$$f(x, y)=x\cdot \pi(y)\oplus g(y)=
x_1(y_1\oplus y_2)\oplus x_2y_1\oplus x_3(y_4\oplus y_5)\oplus x_4(y_3\oplus y_4)\oplus x_5y_3\oplus
y_1y_2y_3y_4y_5\oplus y_2y_3y_4y_5.$$
The linear transformation matrix $A$ is equal to
\begin{eqnarray*}
A=\left(
\begin{array}{cc}
S_5\oplus I_5 & S_5 \\
S_5 & S_5\oplus I_5 \\
\end{array}\right),\ \ {\rm where}
\ \
S_5=\left(
\begin{array}{ccccc}
 0 & 0 & 0 & 0 & 0\\
 1 & 0 & 0 & 0 & 0\\
 1 & 1 & 0 & 0 & 0\\
 1 & 1 & 1 & 0 & 0\\
 1 & 1 & 1 & 1 & 0
\end{array}\right).
\end{eqnarray*}
Therefore, the function $ f'(x, y)=f((x, y)A)=(x_2\oplus
y_1)(x_3x_4x_5y_5\oplus x_3x_4y_4y_5\oplus x_3x_5y_3y_5\oplus
x_3y_3y_4y_5 \oplus x_4x_5y_2y_5\oplus x_4y_2y_4y_5\oplus
x_5y_2y_3y_5\oplus y_2y_3y_4y_5\oplus x_3x_4y_5\oplus
x_3x_5y_5\oplus x_3y_3y_5\oplus x_3y_4y_5\oplus x_4x_5y_5\oplus
x_4y_2y_5\oplus x_4y_4y_5\oplus x_5y_2y_5\oplus x_5y_3y_5\oplus
y_2y_3y_5\oplus y_2y_4y_5\oplus y_3y_4y_5\oplus x_3y_5\oplus
x_4y_5\oplus x_5y_5\oplus y_2y_5\oplus y_3y_5\oplus y_4y_5)\oplus
x_1x_2\oplus x_1y_1\oplus x_2x_3\oplus x_2x_4\oplus x_2x_5\oplus
x_2y_3\oplus x_2y_4\oplus x_3x_5\oplus x_3y_2\oplus x_3y_4\oplus
x_4x_5\oplus x_4y_2\oplus x_4y_3\oplus x_4y_4\oplus x_4y_5\oplus
x_5y_2\oplus x_5y_4\oplus y_1y_2\oplus y_1y_5\oplus y_2y_3\oplus
y_2y_4\oplus y_2y_5\oplus y_3y_5\oplus y_4y_5\oplus x_3\oplus
x_5\oplus y_5$ is bent-negabent and  $\mathrm{deg}(f')=5$.
%
%Computations show that $f'$ is bent-negabent.
\end{Example}

\end{document}